\documentclass[sigconf]{acmart}
\usepackage{multirow}
\usepackage{graphicx}
\usepackage{subcaption}

\AtBeginDocument{%
  }

\renewcommand\footnotetextcopyrightpermission[1]{}
\begin{document}

\title{APG-MOS: Auditory Perception Guided-MOS Predictor for Synthetic Speech}

\author{Zhicheng Lian}
\affiliation{%
  \institution{Beijing Normal University}
  \city{Beijing}
  \country{China}}
\email{zhichenglian@mail.bnu.edu.cn}

\author{Lizhi Wang}
\affiliation{%
  \institution{Beijing Normal University}
  \city{Beijing}
  \country{China}}
\email{wanglizhi@bnu.edu.cn}

\author{Hua Huang}
\affiliation{%
  \institution{Beijing Normal University}
  \city{Beijing}
  \country{China}}
\email{huahuang@bnu.edu.cn}

\begin{abstract}

Automatic speech quality assessment aims to quantify subjective human perception of speech through computational models to reduce the need for labor-consuming manual evaluations. While models based on deep learning have achieved progress in predicting mean opinion scores (MOS) to assess synthetic speech, the neglect of fundamental auditory perception mechanisms limits consistency with human judgments. To address this issue, we propose an auditory perception guided-MOS prediction model (APG-MOS) that synergistically integrates auditory modeling with semantic analysis to enhance consistency with human judgments. Specifically, we first design a perceptual module, grounded in biological auditory mechanisms, to simulate cochlear functions, which encodes acoustic signals into biologically aligned electrochemical representations. Secondly, we propose a residual vector quantization (RVQ)-based semantic distortion modeling method to quantify the degradation of speech quality at the semantic level. Finally, we design a residual cross-attention architecture, coupled with a progressive learning strategy, to enable multimodal fusion of encoded electrochemical signals and semantic representations. Experiments demonstrate that APG-MOS achieves superior performance on two primary benchmarks. Our code and checkpoint will be available on a public repository upon publication.
\end{abstract}

\maketitle

\section{Introduction}

\begin{figure}[t]
    \centering 
    \includegraphics[width=8.5cm]{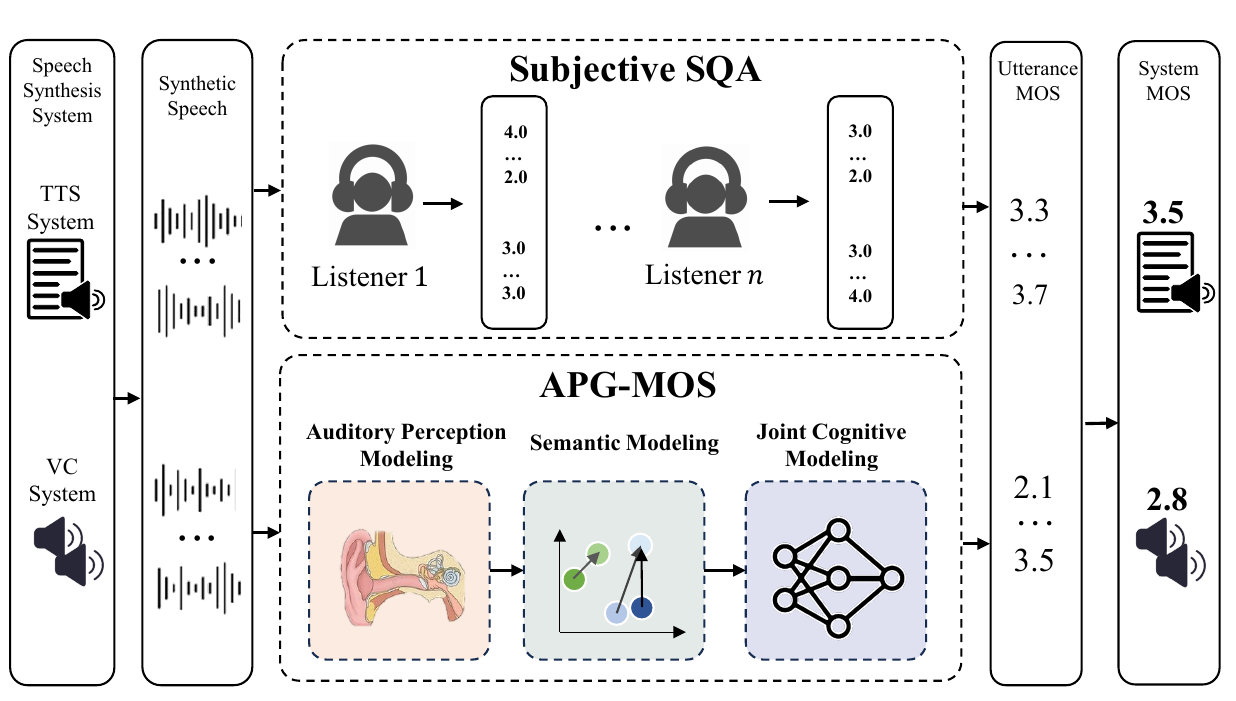} 
    \caption{Illustration of the SQA Task. MOS is a widely adopted metric to assess speech quality for speech synthesis systems such as TTS and VC systems. The subjective listening test for obtaining MOS is a labor-intensive process where each synthetic speech from each system is scored by multiple listeners, whereas automatic SQA can efficiently predict MOS through a computational model. Our proposed APG-MOS is an automatic SQA model which aims to approximate human perception by jointly leveraging auditory perception modeling and semantic modeling.}
    \label{fig:task} 
\end{figure}

Speech quality assessment (SQA) is a critical task in measuring the perceptual quality of speech signals and has broad application in speech communication \cite{WatsonS1998SQA, Fu2018quality, Liu2025Non-intrusive, Mittag2019NISQA, Rix2001PESQ, Beerends2013POLQA} and speech content production \cite{Wang2024TTS, Sheng2023VC, Zhou2024MOS-FAD, Mittag2020NISQA, zhang2024from, Ye2024TTS}. With rapid advancements in speech synthesis technologies, such as text-to-speech (TTS) \cite{Ye2023TTS, Liu2023TTS, Zhu2024TTS, Xiao2024TTS, Huang2022TTS, Liu2024TTS, Deng2023TTS} and voice conversion (VC) \cite{Lu2021VC, Deng2023VC, Sheng2023VC, Ren2023VC}, the diversity and complexity of these systems have increased significantly, which imposes greater demands on SQA methodologies. As shown in Figure \ref{fig:task}, the mean opinion score (MOS) quantifies the quality of a speech sample by averaging subjective ratings from multiple listeners and serves as a standard for assessing synthetic speech quality \cite{cooper2024review}. However, the subjective listening test for obtaining MOS is labor-intensive and time-consuming. As a result, automatic MOS prediction approaches, which aim to approximate subjective human judgments through computational methods, have gained significant attention for their efficiency in assessing synthetic speech quality.

Current SQA methodologies can be broadly categorized into two paradigms: expert knowledge-driven approaches \cite{Rix2001PESQ, Beerends2013POLQA, Malfait2006P.563} and data-driven approaches \cite{Lo2019mosnet, Cooper2022Generalization, saeki2022utmos, Ragano2024SCOREQ}. Expert knowledge-driven approaches rely on biologically inspired auditory models to derive quality-aware characteristics. However, these methods often struggle to model nonlinear processes, limiting the ability to capture semantic-level quality degradation. Data-driven approaches leverage deep neural networks to establish nonlinear mappings between speech signals and subjective quality scores. By incorporating self-supervised learning (SSL) techniques, these approaches enhance the capacity for high-level semantic modeling. However, data-driven approaches often overlook fundamental auditory perception mechanisms, thereby neglecting key processes inherent to human auditory pathways. As a result, both expert knowledge and data-driven approaches exhibit limitations in achieving consistency with human judgments.

To address the challenge, we propose an auditory perception-guided MOS predictor (APG-MOS) for synthetic speech. APG-MOS integrates biologically inspired auditory modeling with advanced semantic analysis, bridging the gap between expert knowledge-driven and data-driven approaches. We simulate the mechanoelectric transduction functions of the cochlea system to encode acoustic signals into biologically aligned electrochemical representations, thereby capturing key auditory information. Then, we propose a residual vector quantization (RVQ)-based method to construct a latent space for semantic distortion, quantifying deviations between raw and quantized SSL representations. Finally, we develop a multilayer residual cross-attention architecture to fuse auditory and semantic features, which employs encoded electrochemical signals to guide the weighting of deep semantic representations. We further optimize the fusion process through a progressive learning strategy. The key contributions of this work are summarized as follows:
\begin{itemize}
\item We simulate the mechanoelectric transduction functions of the cochlea system and introduce biologically aligned auditory representations into deep learning frameworks.
\item We utilize RVQ to formalize a latent space for semantic distortion by capturing residual deviations in SSL representations.
\item We develop a progressive residual cross-attention module for enabling auditory representation to guide semantic representation weighting and propose APG-MOS, achieving superior performance on two primary benchmarks.
\end{itemize}

\section{Related Work}

\textbf{Expert Knowledge-Driven SQA.} Early traditional approaches are based on human auditory characteristics from expert knowledge, which prioritize biologically plausible mechanisms to translate objective signals into quality scores. The classical quality assessment methods PESQ \cite{Rix2001PESQ} and POLQA \cite{Beerends2013POLQA} both employ auditory perception modeling, based on critical band partitioning and auditory masking effects, combined with frequency domain analysis to convert signal distortions into perceptual quality assessments. However, this design inherently relies on reference anchors. The ITU-T P.563 \cite{Malfait2006P.563} is a reference-free algorithm that autonomously evaluates speech quality by hierarchically analyzing distortion characteristics in degraded signals to predict subjective quality scores. Although these methods offer physical interpretability, their reliance on hand-crafted features limits their ability to model the nonlinear relationship between semantic information and speech quality effectively.

\textbf{Data-Driven SQA.} The rise of deep learning has changed the focus to end-to-end MOS prediction. AutoMOS \cite{Patton2016AutoMOS} pioneers waveform-to-MOS mapping using recurrent neural networks. MOSNet \cite{Lo2019mosnet} then integrates convolutional layers to capture spectral-temporal patterns to achieve higher correlation. To address individual listener bias challenges, MBNet \cite{Leng2021MBNet} uses MeanNet to predict average scores and BiasNet to capture listener-specific biases, enhancing the prediction accuracy for utterance-level quality scores. Later work LDNet \cite{Huang2022LDnet} explicitly models listener identities and addresses parameter redundancy in the inference phase of MBNet via an encoder-decoder architecture. Despite listener modeling mitigating human bias in the evaluation of individual utterances, the above methods still struggle to achieve good generalization performance at the system level.

To obtain a higher capability of representations in speech, recent work \cite{Tseng2021utilizing, Cooper2022Generalization, huang2024MOS-Bench} leverages SSL models \cite{Baevski2020wav2vec, Hsu2021HuBERT} pre-trained on large-scale speech datasets, and improves generalization performance. UTMOS \cite{saeki2022utmos} is a system based on ensemble learning that combines strong learners based on the SSL model with weak learners to achieve high-performance MOS prediction. DDOS \cite{tseng2022DDOS} utilizes domain-adaptive pre-training to further pre-train SSL models on synthetic speech for better generalization abilities. Fusion-SSL \cite{yang2022fusion} further enhances cross-domain generalization by combining seven SSL models with semi-supervised learning. Based on this, MOSPC \cite{wang2023mospc} strengthens the ranking capabilities by introducing a pairwise comparison framework. Moreover, some studies use richer information representation, such as spectral \cite{Zezario2023MOSA-Net, baba2024utmosv2} or linguistic characteristics \cite{Voini2023Investigating} as a complement. Recent approaches further integrate technologies such as non-matching reference \cite{manocha2021noresqa}, retrieval-augmentation \cite{wang2023RAMP}, and auditory large language models \cite{Wang2025Enabling}. However, these SSL model-based methods often treat speech as a sequence of vectors without explicitly modeling auditory processes, leaving the fundamental auditory perception mechanisms unaddressed. Our work addresses the limitations of previous work and proposes APG-MOS for this challenge.

\section{Proposed Method}

APG-MOS includes three synergistic components: (1) auditory perception modeling; (2) semantic modeling; and (3) joint cognitive modeling.

\begin{figure*}[t]
    \centering 
    \includegraphics[width=\textwidth]{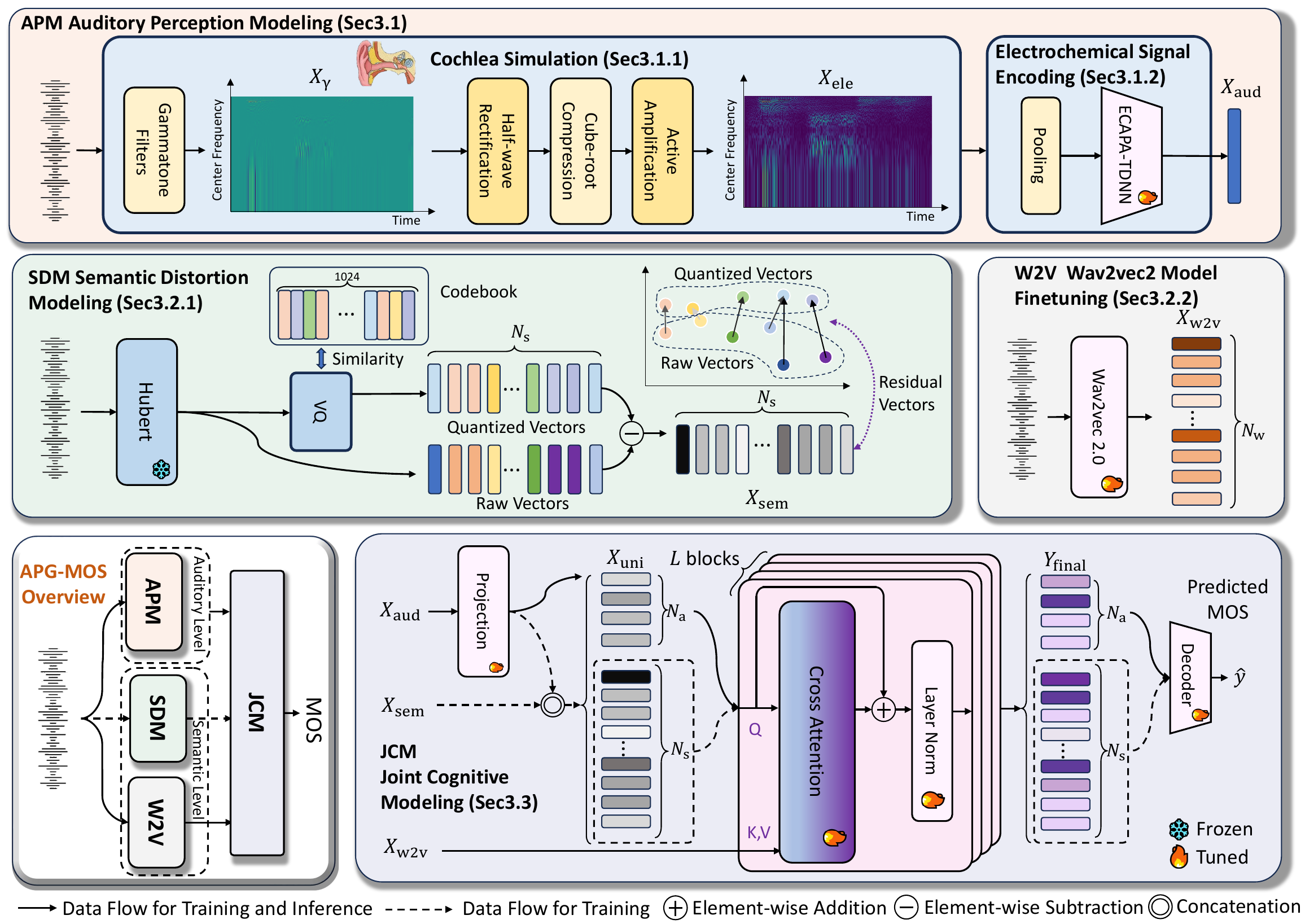} 
    \caption{The main architecture of the proposed APG-MOS method. In the frst stage, auditory perception modeling is proposed to simulate cochlear mechano-electric transduction to encode speech signals into electrochemical representations (Section \ref{sec:auditory}). In the second stage, semantic modeling method is proposed by leveraging HuBERT-RVQ and fine-tuning Wav2vec2 model (Section \ref{sec:semantic}). In the final stage, joint cognitive modeling is proposed to integrate auditory and semantic representations via a residual cross-attention architecture (Section \ref{sec:cognitive}).} %
    \label{fig:overview}
\end{figure*}

\subsection{Auditory Perception Modeling}
\label{sec:auditory}

Simulating the ability of the auditory system to convert acoustic signals into electrochemical signals has long been one of the fundamental paradigms for computational modeling in audio processing. Despite significant advances in speech quality assessment, existing methods often overlook the coding mechanisms that underlie cochlear transduction, limiting their physiological plausibility. We introduce a novel computational framework that integrates biophysical insights from cochlea functions with deep learning architectures. By emulating the tristage signal processing of the basilar membrane spectral decomposition, active amplification, and mechanoelectrical transduction, we establish a physiologically interpretable mapping to transform the raw audio to electrochemical representations. In addition, a compression network with a bottleneck layer is used to obtain a global representation of the auditory perceptual features of speech. Finally, a compact feature representation is generated that conforms to the global characteristics of auditory perception.

\subsubsection{Cochlea Simulation}

Given the pivotal role of the cochlea in the peripheral auditory system, our proposed auditory model emulates the three-stage signal transduction pathway observed in cochlear physiology. This biomimetic framework encompasses: (1) frequency decomposition through basilar membrane mechanics, (2) active amplification governed by outer hair cell motility, and (3) nonlinear mechanoelectrical transduction mediated by inner hair cell stereocilia. As delineated in Figure \ref{fig:overview}, the hierarchical processing follows physiologically grounded signal transformation stages.

Primary to the auditory modeling, the equivalent rectangular bandwidth (ERB) scale \cite{GLASBERG1990ERB} is implemented to establish alignment with the cochlear base-to-apex spatial frequency gradient. Across the human hearing range (20 Hz–20 kHz), $D_f$ center frequencies are non-uniformly spaced according to the ERB function:
\begin{equation}
    \text{ERB}(f) = 24.7 + \frac{f}{9.26449}, 
    \label{eq:erb}
\end{equation}
where the ERB value at frequency $f$ represents the effective bandwidth of the auditory filter, the constant term 24.7 Hz governs apical low-frequency resolution, while the frequency-dependent term $f/9.26449$ captures the stiffness-modulated spectral analysis of the basal region.

Then gammatone filters \cite{patterson1992gammatone} are implemented to simulate the propagation of traveling waves and the frequency selectivity characteristics of the basilar membrane. The impulse response of each filter is defined as:
\begin{equation}
\mathrm{IR}(t) = t^{3} \exp(-2 \pi b \mathrm{ERB}(f) t) \cos(2 \pi f t),
\end{equation}
where $t$ denotes time, $f$ represents the center frequency of the filter (in Hz). The parameter $b$ serves as a bandwidth scaling factor fixed at 1.019 to align with physiological measurements. Each filter exhibits an exponentially decaying temporal envelope and bandpass filtering in the frequency domain, accurately replicating basilar membrane vibration patterns. The filterbank collectively constitutes the passive mechanical filtering stage of cochlear processing.

Next, half-wave rectification is incorporated to model the directional sensitivity of inner hair cell stereocilia deflection \cite{Meddis1986Simulation}. Specifically, half-wave rectification biophysically encodes the unidirectional channel gating mechanism, which is expressed as: 
\begin{equation}
    X_{\text{rec}} = \max(0, X_{\gamma}),
    \label{eq:rectification}
\end{equation}
where $X_{\gamma} \in \mathbb{R}^{N \times D_f}$ denotes the output Gammatone filters, and the parameter $N$ denotes the number of audio time samples. The upward movement of the basilar membrane triggers the influx of potassium ions through the open transduction channels, while the downward movement closes these channels.

Finally, cube-root compression is applied to emulate the nonlinear mechanical-to-electrical transduction in inner hair cells, compressing wide-dynamic-range acoustic signals into neural firing rates. A 3 times gain factor compensates for signal attenuation arising from omitted outer hair cell active amplification mechanisms in the current implementation. The compression process is mathematically expressed as:
\begin{equation}
X_\text{ele} = 3 \cdot (X_{\text{rec}})^{1/3} \quad
\label{eq:compression}
\end{equation}
By this process, we yield a cochleagram, which is a non-negative electrochemical representation $X_\text{ele}\in\mathbb{R}_{\ge0}^{N \times D_f}$ that characterizes the frequency-specific electrochemical responses of inner hair cells, which reflect the spectral and temporal features of acoustic stimuli.

By simulating mechanical signal processing in the cochlea, an auditory-aligned bioelectric signal modality is obtained. Nevertheless, electrochemical signals often include redundant data, necessitating the compression and extraction of perceptual features.

\subsubsection{Electrochemical Signal Encoding}

In order to obtain global features related to speech quality from human auditory perception, we adopt a feature compression network with a bottleneck layer. Bottleneck layers can reduce redundancy by enforcing dimensional compression and maximizing task-relevant information while minimizing irrelevant details.

Specifically, we implement multi-scale temporal pooling to systematically compress the electrochemical representations. First, temporal pooling is applied to downsample the original signal from 16 kHz to 40 Hz by an adaptive average pooling module. This dimensionality reduction strategy retains critical auditory information while reducing the data volume. Then the ECAPA-TDNN model \cite{desplanques2020ecapa} is employed to extract auditory perception features from simulated cochlear processed electrochemical signals, which is expressed as: 
\begin{equation}
X_{\text{aud}} = \text{ECAPA-TDNN}(\text{Pooling}(X_{\text{ele}})),
\end{equation}
where the model takes the electrochemical signal gram $X_\text{ele}$ as input. The model processes input features through a TDNN module to capture temporal local information and compresses temporal features into a global representation using attentive statistical pooling. Finally, the model ends with a bottleneck of dimension $D_a=192$ using a fully connected layer and a batch normalization layer, which outputs a compact global feature $X_{\text{aud}}\in\mathbb{R}^{D_a}$ encoding human auditory perception.

In general, we present a bioinspired computational model that creates the cochlear signal transduction pipeline aligned to biological mechanisms. By incorporating ERB-scale Gammatone filtering, half-wave rectification, active amplification, and cube-root compression, the model captures the frequency selectivity, directional sensitivity, and dynamic compression observed in human hearing. By integrating biophysical modules with the ECAPA-TDNN architecture, the model further enables the extraction of discriminative temporal and spectral features.

\subsection{Semantic Modeling}

\label{sec:semantic}

Semantic-level representations guide the selection of appropriate acoustic parameters that convey the intended meaning. Recent methods often implicitly construct semantic information by fine-tuning the pre-trained SSL-model. With the adoption of the fine-tuning strategy, we further explicitly model the semantic distortion space utilizing RVQ technology.

\subsubsection{Semantic Distortion Modeling}

HuBERT\cite{Hsu2021HuBERT} is a Transformer-based self-supervised speech representation learning model that effectively captures phonetic, lexical, and other semantic information, providing a strong semantic prior for downstream tasks. The residual vector quantization is a hierarchical architecture that consists of stacked vector quantization (VQ) layers, each incorporating a Euclidean Codebook module capturing residual errors through iterative refinement. The forward pass process can be expressed as $q_i = \text{VQ}_i(r_{i-1})$ where $q_i$ represents the quantized vectors of $\text{VQ}_i$, and $r_i$ is the residual vector after $\text{VQ}_i$, calculated as \(r_i=r_{i - 1}-q_i\) with \(r_0\) being the first input vectors. In this work, the codebook of the first VQ layer \(\text{VQ}_1\) is adopted since the discrete semantic vectors extracted by this layer mainly correspond to phonetic units capturing features of phones and timbre. Next, the residual vectors between the raw HuBERT and the quantized vectors of \(\text{VQ}_1\) are given by:
\begin{equation}
X_\text{sem} = X_\text{h} - \text{VQ}_1(X_\text{h}),
\end{equation}
where $X_\text{h} = \text{HuBERT}(x)$ represents the raw vectors encoded by the HuBERT model. The residual deviations are represented by $X_{\text{sem}}\in\mathbb{R}^{N_\text{s} \times D_\text{s}}$, which formalizes a latent space for semantic distortion.

To explain this, the initial VQ layer captures the most representative and normal phonetic units, having pre-trained on clean speech data. As a result, $X_\text{sem}$ signifies the information that the initial VQ layer does not capture, indicating its deviation from the typical units at the semantic level. When the deviations expressed by \(X_{\text{sem}}\) are slight, it means that the speech is better aligned with the codebook. This alignment implies that the speech not only conveys the correct semantic information, but also presents it in a more natural and high-quality way. In contrast, if the deviations are substantial, it indicates a misalignment between speech and semantic codebooks. This misalignment shows a gap from the standard phonic units and implies the unnaturalness of the speech.

\subsubsection{Wav2vec2 Model Fine-tuning}

To take advantage of the implicit semantic modeling capabilities of self-supervised models, the Wav2vec2 model \cite{Baevski2020wav2vec} is incorporated. Deep embedding $X_{\text{w2v}}\in\mathbb{R}^{N_\text{w} \times D_\text{w}}$ is obtained by fine-tuning MOS prediction datasets with a connection to a decoder module. The wav2vec2 model is pre-trained on a large-scale unlabeled speech dataset to learn rich speech feature representations. The pre-trained wav2vec2 model can capture high-level semantic information in speech, such as speech rhythm and linguistic information, and is widely used for speech quality-related tasks to provide a richer speech context \cite{xie2024An, Cooper2022Generalization}.

\subsection{Joint Cognitive Modeling}
\label{sec:cognitive}

Although the above methods significantly narrow the gap between speech and its quality from the human auditory perception and semantic distortion perspective, there remain multimodal challenges for constructing a cognitive pathway from auditory electrochemical encodings to semantic understanding and finally to speech quality assessment. To address the challenge, we propose a residual architecture based on cross-attention with a progressive training strategy that constructs a hierarchical cognitive pathway to fuse multisensory features, thus enhancing the discriminative capacity of speech quality through multimodal interactions.

\subsubsection{Modal Fusion Architechture}

To align auditory and semantic features, auditory perception encoding vectors $X_\text{aud}$ are first mapped through a trainable projection layer to produce a fixed length $N_a=8$ vector sequence with dimension $D_{\text{s}}$ of each vector that matches the dimension of $X_{\text{sem}}$. Subsequently, these two vector sequences are concatenated along the dimension of the audio frames (as shown in Figure \ref{fig:overview}), creating a unified latent space that projects auditory features into the semantic distortion space to form joint embeddings $X_{\text{uni}}$, which can be expressed as:
\begin{equation}
X_{\text{uni}} = [\text{Linear}(X_\text{aud}), X_{\text{sem}}]
\end{equation}
This cross-modal alignment strategy aims to preserve details from the auditory domain and bridge the gap between multiple modalities.

Subsequently, the connection between the joint embedding and the wav2vec2-based representation is achieved through a module consisting of $L$-layer cross-attention modules with residual connections. Within the cross-attention module, the embedding \(X_{\text{uni}}\) acts as the query sequence, and the feature $X_{\text{w2v}}$ extracted by the wav2vec2 model acts as both key and value matrices. This can be expressed as:
\begin{equation}
Q^{l}=X^{l}W_{Q}^{l}, K^{l}=X_{\text{w2v}}W_{K}^{l},V^{l}=X_{\text{w2v}}W_{V}^{l},
\end{equation}
where \(W_{Q}^{l}, W_{K}^{l}, W_{V}^{l}\) are the trainable weight matrices corresponding to the query, key and value projections for the $l$-th layer of the cross-attention modules, with $X^{1} = X_{\text{uni}}$ and \(l = 1,\cdots,L\) for each layer. After calculating the cross-attention results, they are processed by layer normalization to ensure the stability and consistency of the features. Then, the normalized results are residually connected with the results of the previous layer. The process can be represented by the following formula:
\begin{equation}
\begin{cases}
Y^{l}=\text{Norm}(\text{Attention}(Q^{l}, K^{l}, V^{l})+X^{l})\\
X^{l + 1}=Y^{l}
\end{cases}
\end{equation}
where the final result $Y^{\text{fusion}}=Y^{L}$. In this way, the cross-attention mechanism allows the model to selectively guide and weight relevant information from the high-level semantic features based on the information of human auditory perception. The introduction of residual connections can effectively alleviate the gradient vanishing problem, enabling the model to more easily learn the feature differences between different layers, thereby improving the training efficiency and performance of the model.

\begin{figure}[t]
    \centering 
    \includegraphics[width=8.5cm]{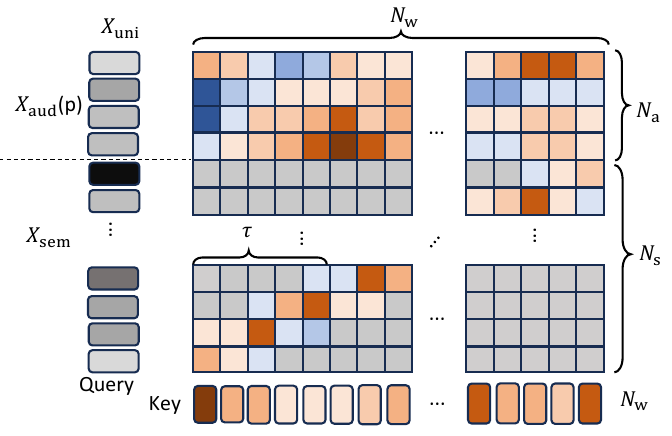} 
    \caption{Illustration of the attention mask.} 
    \label{fig:mask} 
\end{figure}

In addition, the attention mechanism employs specialized masking strategies to handle feature interactions. For projected $X_\text{uni}$ embeddings, no mask is applied to preserve global contextual relationships. In contrast, the embedding of semantic distortion $X_{\text{sem}}$ uses a masking scheme in which a diagonal band mask with a frame width is applied to the attention score matrix, restricting each query to focus only on neighboring key frames within in-context frames (as shown in Figure \ref{fig:mask}). This design enables the model to jointly learn the corresponding relations while addressing the inherent temporal alignment between features with different sampling rates. The diagonal band mask is formally defined as:  
\begin{equation}
M_{ij} = \begin{cases} 
0, & |\lambda \cdot (i-N_{\text{a}})-j| > \tau \ \text{and} \ i \ge N_{\text{a}}\\
1, & (|\lambda \cdot (i-N_{\text{a}})-j| \leq \tau \ \text{and} \ i \ge N_{\text{a}}) \ \text{or} \ i<N_{\text{a}}
\end{cases}
\end{equation}
where $i$ and $j$ are respectively indexes for feature sequence $X_\text{uni}$ and $X_\text{w2v}$. The parameter \( \lambda = N_\text{w} / N_\text{s} \) is a scaling factor to align the temporal resolutions between $X_\text{sem}$ and $X_{\text{w2v}}$. And \( \tau = 10 \) is a bandwidth threshold that controls the local attention range, restricting each key to respond only to queries within in-context frames. This balances local granularity with global alignment, mitigating temporal misalignment while preserving critical correlations.

After being processed by the residual cross-attention module, the obtained results are fed into a decoder network to calculate the final MOS prediction $\hat{y}$ of the speech, which is given by:
\begin{equation}
\hat{y} = \text{Scalar}(\tanh(\text{Linear}(\text{ReLU}(\text{Linear}(Y^{\text{fusion}})))),
\end{equation}
where the $\tanh$ activation function maps the output results to the range of \((-1, 1)\), and the linear offset scaler module calculates the mean of the results and normalizes them to the range of \((1, 5)\), which is in line with the common range of MOS scores.

\subsubsection{Progressive Learning Strategy}

To address the overfitting issue caused by limited training data in complex models, a progressive learning strategy is designed, where different modules are integrated through a phased training approach and the results are optimized through feature pruning during the inference phase.

\begin{figure}[t]
    \centering 
    \includegraphics[width=8.5cm]{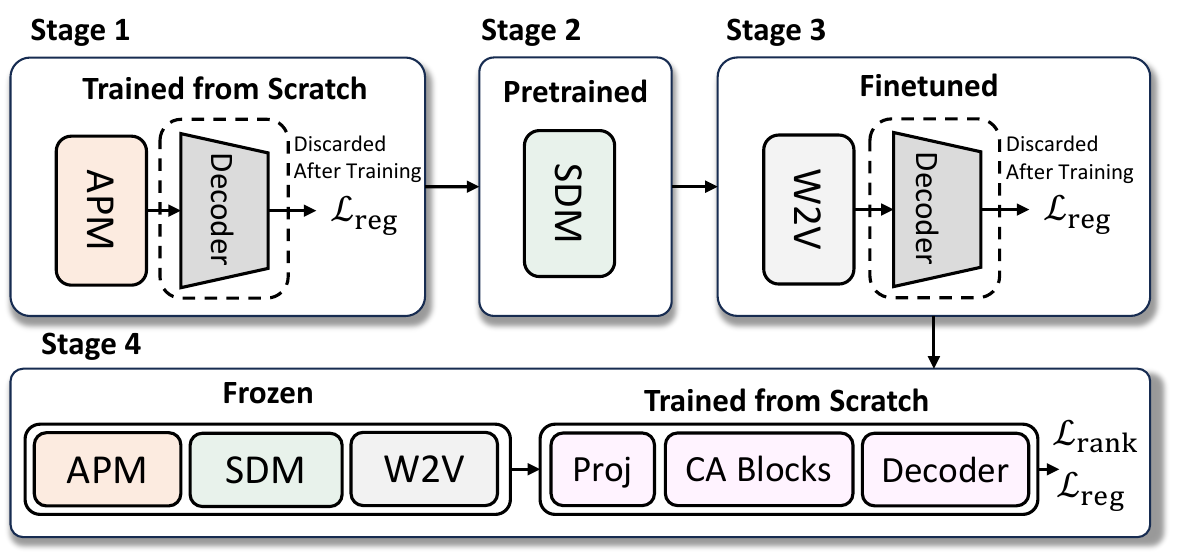} 
    \caption{Illustration of the progressive training strategy.} 
    \label{fig:training} 
\end{figure}

As indicated in Figure \ref{fig:training}, the auditory perception modeling module (APM), the semantic distortion modeling module (SDM) and the Wav2vec2 (W2V) employ distinct training strategies. Firstly, the APM module is trained from scratch on a MOS-annotated speech dataset by feeding the encoded embeddings $X_{\text{ele}}$ to the decoder. In the SDM module, HuBERT and RVQ are both pre-trained on unlabeled clean speech datasets (specific details can be referenced in work\footnotemark[1]). The W2V model undergoes parameter initialization using pre-trained weights followed by task-specific fine-tuning on MOS prediction data. Upon convergence of individual modules, all learned parameters in APM, SDM, and W2V are fixed to preserve their acquired feature representations. The downstream architecture components, including cross-attention blocks, projection layer, and decoder, are optimized using the MOS dataset.

\footnotetext[1]{\url{https://github.com/RVC-Boss/GPT-SoVITS}}

During inference, the process is partially decoupled from the training process. In the training process, the SDM module serves two main purposes. On the one hand, SDM leverages the advantages of being trained on clean datasets to avoid overfitting to smaller task-specific datasets. On the other hand, SDM constructs a generalized latent space that directs auditory features to project into a consistent embedding, reducing the representation gap between auditory and semantic levels. However, during inference, features trained without task-specific data supervision may inadvertently introduce noise. Therefore, the features extracted by SDM are systematically pruned during the inference process, retaining only $X_{\text{uni}} = [\text{Linear}(X_\text{aud})]$ for the computation of cross-attention. This strategy achieves better performance compared to the full-feature model while reducing computational complexity.

For loss functions, we adopt the L1 regression loss \(\mathcal{L}_{\text{reg}}\) to train APM and W2V. And we use \(\mathcal{L}_{\text{total}}\), which harmonizes the relative ranking loss \(\mathcal{L}_{\text{rank}}\) and the L1 regression loss \(\mathcal{L}_{\text{reg}}\) to train the modules in joint cognitive modeling (as shown in Figure \ref{fig:training}). The formula is given by:
\begin{equation}
\mathcal{L}_{\text{total}}=(1 - \alpha)\cdot \mathcal{L}_{\text{rank}}+\alpha\cdot \mathcal{L}_{\text{reg}},
\end{equation}
where \(\alpha\) is a hyperparameter used to balance the weights of the relative ranking loss and the L1 losses. The detailed description of \(\mathcal{L}_{\text{rank}}\) and \(\mathcal{L}_{\text{reg}}\) refers to the supplementary material.

\begin{table*}[t]
\centering
\caption{Performance comparison of system-level results with different models on BVCC and SOMOS Datasets.}
\label{tab:model_comparison}
\renewcommand{\arraystretch}{1.0}
\setlength{\tabcolsep}{8.9pt}
\begin{tabular}{@{}lcccccccccc@{}}
\toprule
\multirow{2}{*}{Model} & \multirow{2}{*}{\shortstack{Model\\Base}} & \multirow{2}{*}{\shortstack{Prior\\Dependency}} & 
\multicolumn{4}{c}{BVCC} & \multicolumn{4}{c}{SOMOS} \\
\cmidrule(lr){4-7} \cmidrule(lr){8-11}
 & & & {MSE$\downarrow$} & {LCC$\uparrow$} & {SRCC$\uparrow$} & {KTAU$\uparrow$} & {MSE$\downarrow$} & {LCC$\uparrow$} & {SRCC$\uparrow$} & {KTAU$\uparrow$} \\
\midrule
MOS-Net\cite{Lo2019mosnet}         & - & -   & 0.563 & 0.261 & 0.266 & --    & 0.081 & 0.667 & 0.679 & 0.479 \\
MBNet\cite{Leng2021MBNet}           & - & L   & 0.228 & 0.844 & 0.870 & 0.685 & \multicolumn{4}{c}{--} \\
LDNet\cite{Huang2022LDnet}           & - & L   & 0.139 & 0.896 & 0.893 & 0.714 & \textbf{0.045} & 0.849 & 0.847 & 0.639 \\
MOSA-Net\cite{Zezario2023MOSA-Net}        & W,H & -   & 0.162 & 0.899 & 0.900 & 0.729 & \multicolumn{4}{c}{--} \\
SSL-MOS\cite{Cooper2022Generalization}         & W & -   & 0.184 & 0.928 & 0.928 & 0.776 & 0.052 & 0.911 & 0.918 & 0.741 \\
RAMP(P)\cite{wang2023RAMP}         & W & -   & \textbf{0.106} & 0.917 & 0.915 & 0.756 & 0.182 & 0.664 & 0.659 & 0.474 \\
Vioni et al.\cite{Voini2023Investigating}    & W,B & L,T   & \multicolumn{4}{c}{--} & 0.052 & 0.911 & 0.917 & 0.741 \\
DDOS\cite{tseng2022DDOS}            & W & L   & 0.110 & 0.933 & 0.932 & 0.782 & \multicolumn{4}{c}{--} \\
UTMOS(strong)\cite{saeki2022utmos}   & W & L,P & 0.148 & 0.930 & 0.925 & 0.774 & \multicolumn{4}{c}{--} \\
\textbf{APG-MOS}       & W & -   & 0.120 & \textbf{0.934} & \textbf{0.936} & \textbf{0.784} & 0.072 & \textbf{0.916} & \textbf{0.921} & \textbf{0.746} \\
\bottomrule
\end{tabular}
\vskip 0.1cm
\small
\textit{Note:} W=Wav2vec2(base), B=Bert(base), H=HuBERT(base), L=Listener Label, P=Phoneme Label, T=Text Label.
\end{table*}

\section{Experiment}

\subsection{Implementation Details}
\subsubsection{Datasets} Our experiments are conducted on the BVCC\cite{Cooper2022Generalization} (Blizzard Challenge and Voice Conversion Challenge combined dataset) and SOMOS\cite{maniati2022somos} (a high-quality speech dataset focusing on neural speech synthesis systems). BVCC contains 7,106 English synthetic speech samples covering 187 systems, while SOMOS comprises 20K synthetic speech samples derived from the LJSpeech dataset with crowdsourced naturalness MOS annotations across 201 systems. We use the official train, dev, and test partitions for both datasets.

\subsubsection{Implementation Details} The hyperparameter \(\alpha\) to balance the weights of relative ranking loss and regression losses is set to 0.9. The model is implemented using Pytorch and trained on a single NVIDIA A800 GPU. The cochlea simulation in the APM module is implemented based on brian2hears\footnotemark[2]. The network is trained using the stochastic gradient descent optimizer (SGD), with a momentum of 0.9 and a learning rate of $0.001$.

\footnotetext[2]{\url{https://brian2hears.readthedocs.io/en/stable/}}

\begin{table*}[t]
\centering
\caption{Performance comparison of system-level results with resource-demanding models on BVCC and SOMOS Datasets.}
\label{tab:model_comparison_source}
\renewcommand{\arraystretch}{1.0}
\setlength{\tabcolsep}{9.25pt}
\begin{tabular}{@{}lcccccccccc@{}}
\toprule
\multirow{2}{*}{Model} & \multirow{2}{*}{\shortstack{Model\\Base}} & \multirow{2}{*}{\shortstack{Prior\\Dependency}} & 
\multicolumn{4}{c}{BVCC} & \multicolumn{4}{c}{SOMOS} \\
\cmidrule(lr){4-7} \cmidrule(lr){8-11}
 & & & {MSE$\downarrow$} & {LCC$\uparrow$} & {SRCC$\uparrow$} & {KTAU$\uparrow$} & {MSE$\downarrow$} & {LCC$\uparrow$} & {SRCC$\uparrow$} & {KTAU$\uparrow$} \\
\midrule
Wang et al.\cite{Wang2025Enabling}      & S(13.75B) & -   & 0.152 & 0.884 & 0.884 & --     & \textbf{0.034} & 0.894 & 0.891 & --     \\
UTMOS\cite{saeki2022utmos}  & C1(1.62B) & L,P & \textbf{0.090} & \textbf{0.939} & \textbf{0.936} & \textbf{0.795} & \multicolumn{4}{c}{--} \\
RAMP\cite{wang2023RAMP}             & W & DS   & 0.097 & 0.931 & 0.932 & 0.782 & \multicolumn{4}{c}{--} \\
RAMP(NP)\cite{wang2023RAMP}         & W & DS   & 0.101 & 0.934 & 0.934 & 0.787 & 0.174 & 0.676 & 0.665 & 0.480 \\
\textbf{APG-MOS}      & W(95M) & -   & 0.120 & 0.934 & \textbf{0.936} & 0.784 & 0.072 & \textbf{0.916} & \textbf{0.921} & \textbf{0.746} \\
\bottomrule
\end{tabular}
\vskip 0.1cm
\small
\textit{Note:} S=SALMONN\cite{tang2024salmonn} (an auditory large language model), L=Listener, P=Phoneme. C1 is a stacking stragegy of 17 strong learners and 6 weak learners using Wav2vec2(base), HuBERT(base), and WavLM(base)\cite{chen2022wavlm} models, DS=DataStore.
\end{table*}

\subsection{Experimental Results}

Evaluation metrics include system-level mean squared error (MSE), linear correlation coefficient (LCC), Spearman rank correlation coefficient (SRCC) and Kendall tau (KTAU) \cite{huang2024MOS-Bench}. For detailed formulas of the metrics, please refer to supplementary material. Here, system-level results refer to aggregating mean MOS scores across all utterances from each speech synthesis system, which assesses the capability of the model to objectively assess the overall quality of speech generation systems. SRCC is the most important metric, as it preserves the relative ordering of system performance that human listeners inherently perceive.

Table \ref{tab:model_comparison} shows the results of the comparative experiment. APG-MOS achieves system-level SRCC scores of 0.936 on BVCC and 0.921 on SOMOS, demonstrating significant superiority over existing models. Compared to approaches such as SSL-MOS, APG-MOS exhibits comprehensive advantages across LCC, SRCC, and KTAU metrics. Supported jointly by auditory perception modeling and semantic modeling, even when benchmarked against models that require listener labels or phoneme annotations (\textit{e.g.} UTMOS (strong), DDOS), APG-MOS maintains the highest SRCC performance on both datasets. The advance lies in the hierarchical modeling of APG-MOS with the auditory-semantic-cognitive processing continuum, where the model prioritizes human-centric systemic cognitive variations across speech synthesis systems rather than modeling listener-specific local quality variations in individual samples.

As demonstrated in Table \ref{tab:model_comparison_source}, APG-MOS yields results on par with its more resource-demanding counterparts. While matching the SRCC performance of UTMOS, our architecture requires only 95M Wav2vec2 parameters plus an additional 14M parameters for feature fusion. Furthermore, APG-MOS surpasses the retrieval-augmented RAMP model by using an external datastore in both SRCC and LCC metrics, particularly excelling on the SOMOS dataset. This performance gain stems from our introduction of the lightweight auditory perception modeling, which constructs a pathway closer to human real perception, rather than mere parameter scaling. For more details of the compared models, please refer to the supplementary material.

\subsection{Ablation Study}

\begin{table*}[t]
\centering
\caption{Ablation Study on BVCC and SOMOS Datasets.}
\label{tab:ablation}
\renewcommand{\arraystretch}{1.0}
\setlength{\tabcolsep}{9.9pt}
\begin{tabular}{@{}lcccccccccc@{}}
\toprule
\multirow{2}{*}{Model} & \multirow{2}{*}{\shortstack{Model\\Base}} & \multirow{2}{*}{\shortstack{Prior\\Dependency}} & 
\multicolumn{4}{c}{BVCC} & \multicolumn{4}{c}{SOMOS} \\
\cmidrule(lr){4-7} \cmidrule(lr){8-11}
 & & & {MSE$\downarrow$} & {LCC$\uparrow$} & {SRCC$\uparrow$} & {KTAU$\uparrow$} & {MSE$\downarrow$} & {LCC$\uparrow$} & {SRCC$\uparrow$} & {KTAU$\uparrow$} \\
\midrule
Mel+D & - & - & 0.264 & 0.829 & 0.825 & 0.639 & 0.059 & 0.726 & 0.722 & 0.529 \\
APM+D   & - & - & 0.163 & 0.874 & 0.872 & 0.691 & 0.061 & 0.743 & 0.747 & 0.545 \\
\hline
QH+D & H & - & 0.640 & 0.645 & 0.653 & 0.469 & 0.111 & 0.450 & 0.433 & 0.294 \\
SDM+D & H & - & 0.427 & 0.846 & 0.858 & 0.667 & 0.109 & 0.585 & 0.581 & 0.410 \\
\hline
w/o APM   & H,W & - & 0.139 & 0.929 & 0.933 & 0.781 & 0.040 & 0.915 & 0.920 & 0.744 \\
w/o SDM   & W & - & \textbf{0.107} & 0.932 & 0.932 & 0.778 & \textbf{0.026} & 0.913 & 0.918 & 0.739 \\
w/o CA    & W & - & 0.136 & 0.931 & 0.931 & 0.778 & 0.052 & 0.915 & 0.919 & 0.743 \\
w/o RL    & W & - & 0.124 & 0.932 & 0.935 & 0.783 & 0.085 & \textbf{0.919} & 0.920 & 0.745 \\
ASI       & H,W & - & 0.110 & 0.931 & 0.933 & 0.780 & 0.076 & 0.917 & 0.920 & \textbf{0.746} \\
SI        & H,W & - & 0.140 & 0.930 & 0.932 & 0.780 & 0.075 & 0.916 & 0.919 & \textbf{0.746} \\
APG-MOS    & W & - & 0.120 & \textbf{0.934} & \textbf{0.936} & \textbf{0.784} & 0.072 & 0.916 & \textbf{0.921} & \textbf{0.746} \\
\bottomrule
\end{tabular}
\vskip 0.1cm
\small
\textit{Note:} APM=Auditory Perception Modeling, SDM=Semantic Distortion Modeling, D=Decoder, ASI=Auditory \& Semantic feaure-based Inference, PI=Semantic feature-based Inference, CA=Cross-Attention, RL=Ranking Loss, W=Wav2vec2(Base), H=HuBERT(base), QH=Quantized HuBERT(base).
\end{table*}

\begin{figure}[h]
    \centering 
    \includegraphics[width=8.5cm]{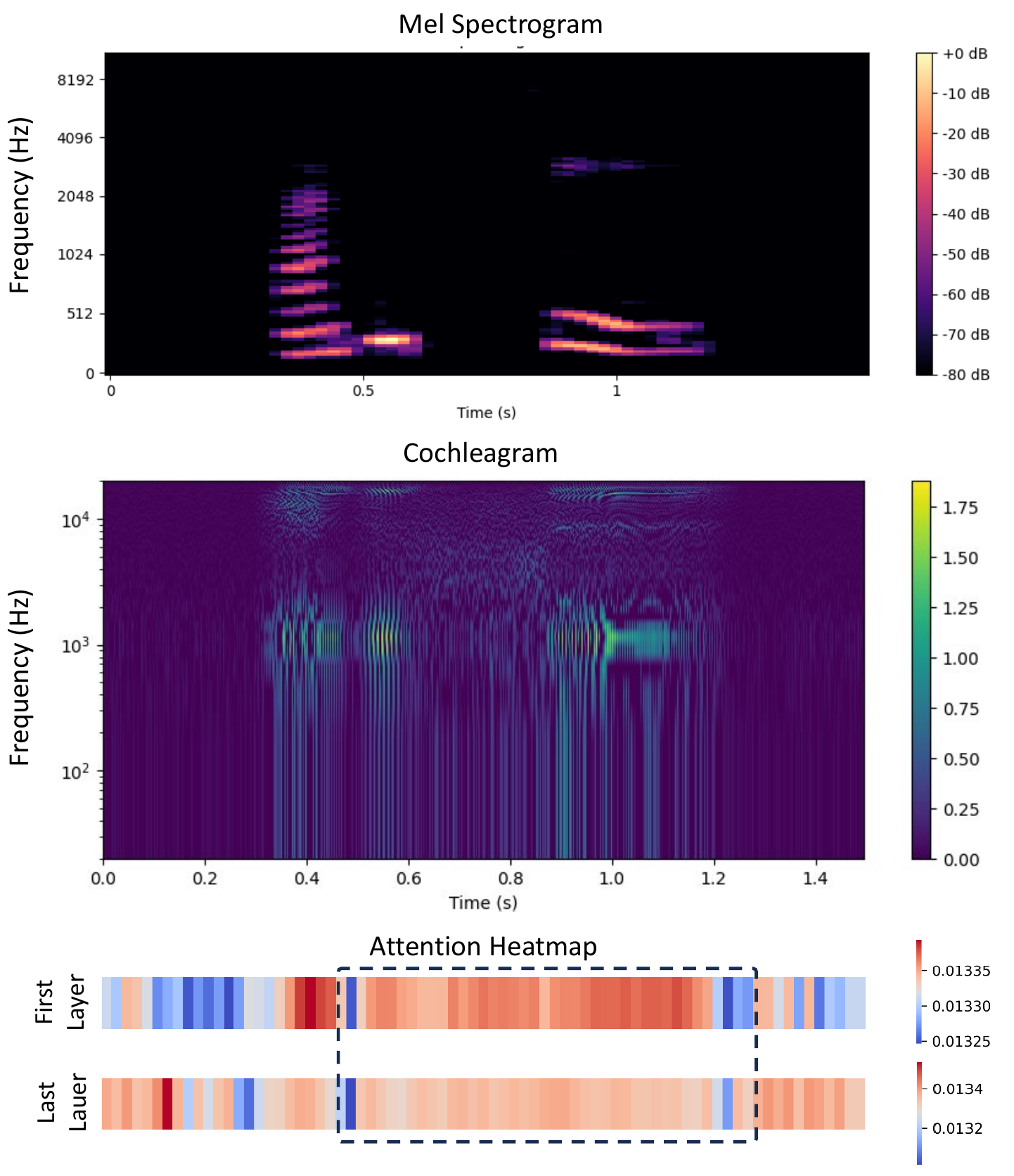} 
    \caption{Visualization of the Mel spectrogram, cochleagram $X_{\text{ele}}$, weight heatmap of cross-attention blocks for the first and the last layer of a sample "sys520c6-utt3f40b05.wav" in BVCC dataset. The part encased in the dashed box can illustrate the impact of auditory features in guiding the weights of SSL model-based features from shallow to deep.} 
    \label{fig:vis} 
\end{figure}

Table \ref{tab:ablation} illustrates the validation of our architectural design through ablation studies. Removing the APM Module while retaining the SDM module degrades SRCC on both datasets, confirming the critical role of APM in supplementing quality prediction. Conversely, preserving APM while removing SDM achieves the lowest MSE on both BVCC and SOMOS, but causes a decrease in SRCC on both datasets. This indicates that although SDM introduces minor noise (MSE increases), SDM mitigates overfitting and enhances the ranking sensitivity by capturing quality degradations at the semantic level (SRCC increases).

To further elucidate the capabilities of APM and SDM, we establish additional baseline comparisons. For the APM module, we implement a Mel spectrogram-based baseline where 128 Mel filter banks are employed, encoded also by an ECAPA-TDNN and decoder module. Experimental results demonstrate that the APM-based approach significantly outperforms the Mel-spectrogram baseline across BVCC and SOMOS datasets, confirming that detailed auditory perception mechanism simulation surpasses simplistic Mel-spectrogram modeling in capturing speech quality characteristics. Visualizations of Mel spectrogram and cochleagram are presented in Figure \ref{fig:vis}. It can be seen that the cochleagram provides a frequency response distinct from that of the Mel spectrogram while retaining more details. For more visualizations of speech samples, please refer to the supplementary material of this paper.

For the SDM module, we compare the vectors extracted by SDM with the quantized HuBERT vectors by connecting to a trainable decoder module with all other parameters frozen. Our SDM-derived semantic distortion features exhibit stable performance improvements across both datasets, achieving substantial gains over HuBERT vectors quantized with VQ. This discrepancy arises because the quantized vectors capture the primary semantic information from clean speech data but are irrelevant to deteriorating speech quality. Conversely, the residual vectors extracted by SDM, while stripped of primary semantic content, precisely quantify deviations from standard phonic units through residual measurements, which proves particularly effective for speech quality evaluation.

Our investigation into inference strategy selection reveals distinct performance patterns when employing different feature types as attention queries under identical training conditions. Experimental results demonstrate a clear hierarchy that APG-only inference outperforms joint inference, which in turn surpasses the SDM-only inference approach. The primary limitation stems from semantic distortion features - while effectively capturing distortion representations in semantic space within clean datasets, SDM inherently introduces quantization noise that compromises stability improvements. However, these residuals prove indispensable during model training, serving critical functions in guiding the projection of auditory features into the semantic space and acting as implicit regularizers. This mechanism facilitates the crucial alignment between auditory characteristics, semantic representations, and ultimate quality assessments.

In addition, as shown in Table \ref{tab:ablation}, the ablation experiment on ranking loss reveals that ranking loss slightly improves SRCC, which marginally enhances the sensitivity of the model to quality ranking. Furthermore, removing the Cross-Attention (CA) module reduces SRCC on BVCC, demonstrating the superior capability of the attention mechanism in fusing auditory and semantic features compared to naive feature concatenation. This degradation highlights the limitations of simple concatenation in fully exploiting cross-modal interactions between the multimodal representations. 

In our multilayer residual attention architecture, the first attention layer receives input auditory and semantic representations independently, while deeper layers establish residual fusion connections between auditory and semantic representations. The visualization of the attention heatmap is shown in \ref{fig:vis}, which reveals how the attention of the first layer guides and influences deeper processing stages, with semantic information progressively refined through residual propagation between layers. From the visualization of this sample, we can observe a significant correlation between the first and last layers of cross-attention, indicating that the auditory representation has a notable impact on the results. For more visualizations of speech samples, please refer to supplementary material. This hierarchical design mimics the staged processing of the auditory system from auditory perception to semantic interpretation.

\section{Conclusion}

In this work, we propose APG-MOS, an auditory perception-guided MOS predictor for automatic synthetic speech quality assessment by synergizing cochlear-inspired auditory modeling with semantic modeling. A biologically aligned auditory module is introduced to encode raw speech signals into electrochemical representations. An RVQ-based semantic distortion modeling method is utilized to characterize quality degradation at the semantic level. A progressive residual cross-attention architecture that hierarchically fuses auditory and semantic features with a progressive learning strategy. Extensive experiments on BVCC and SOMOS datasets demonstrate that APG-MOS achieves superior performance with system-level SRCC of 0.936 and 0.921, respectively. This work underscores the importance of integrating expert knowledge in the auditory physiology domain with advanced deep learning techniques for SQA and establishes a better biologically aligned SQA system.

\bibliographystyle{ACM-Reference-Format}
\bibliography{reference}

\end{document}


\title{Supplementary Material for \\ APG-MOS: Auditory Perception Guided-MOS Predictor for Synthetic Speech}

\author{Zhicheng Lian}
\affiliation{%
  \institution{Beijing Normal University}
  \city{Beijing}
  \country{China}}
\email{zhichenglian@mail.bnu.edu.cn}

\author{Lizhi Wang}
\affiliation{%
  \institution{Beijing Normal University}
  \city{Beijing}
  \country{China}}
\email{wanglizhi@bnu.edu.cn}

\author{Hua Huang}
\affiliation{%
  \institution{Beijing Normal University}
  \city{Beijing}
  \country{China}}
\email{huahuang@bnu.edu.cn}

\maketitle

In the supplementary material, we provide the following. 
\begin{itemize}
    \item In Section \ref{loss}, we detail the formulas of the loss functions.
    \item In Section \ref{eval}, we show the formulas of the evaluation metrics.
    \item In Section \ref{vis}, we visualize the Mel spectrograms, cochleagrams, and attention weight heatmaps of more speech samples.
    \item In Section \ref{compare}, we introduce detailed descriptions of comparative methods in the experiments.
\end{itemize}
\appendix

\section{Details for Loss Functions}
\label{loss}

As a supplement to Section 3.3.2, we detail the two loss functions used. Integrating both numerical regression and relative ranking objectives shows a useful effect in improving model ranking sensitivity \cite{wang2023mospc}. We adopt a loss function \(\mathcal{L}_{\text{total}}\), which harmonizes the mean relative ranking loss \(\mathcal{L}_{\text{rank}}\) and the L1 regression loss \(\mathcal{L}_{\text{reg}}\) at the frame level. This design enables the model not only to predict accurate mean opinion score (MOS) values but also to discern subtle quality differences between speech samples. 

Specifically, the expression of relative ranking loss \(\mathcal{L}_{\text{rank}}\) is:
\begin{equation}
\mathcal{L}_{\text{rank}}=-M\cdot\log(P)-(1 - M)\cdot\log(1 - P),
\end{equation}
where the probability \(P\) is calculated by the logistic function:
\begin{equation}
P=\frac{e^{\hat{y}_{i}-\hat{y}_{j}}}{1 + e^{\hat{y}_{i}-\hat{y}_{j}}}.
\end{equation}
Here, \(\hat{y}_{i}\) and \(\hat{y}_{j}\) represent the predicted MOS scores of the model for any pair of speech samples in a batch, respectively. The value of \(M\) depends on the true MOS scores \(y_i\) and \(y_j\) of the two samples:
\begin{equation}
M = \begin{cases} 
0, & y_i < y_j \\
0.5, & |y_i - y_j| < \beta \\
1, & y_i > y_j 
\end{cases}
\end{equation}
where $\beta$ is set to 0.1 to ensure that the loss function effectively discriminates between samples with non-negligible quality disparities while assigning intermediate values to marginally distinct cases. The L1 losses \(\mathcal{L}_{\text{reg}}\) are calculated as the absolute errors between the predicted MOS scores and the true scores respectively:
\begin{equation}
\mathcal{L}_{\text{reg}}=\vert \hat{y}_{i}-y_i\vert.
\end{equation}

\begin{figure*}[htbp]
    \centering
    \includegraphics[width=1.0\textwidth]{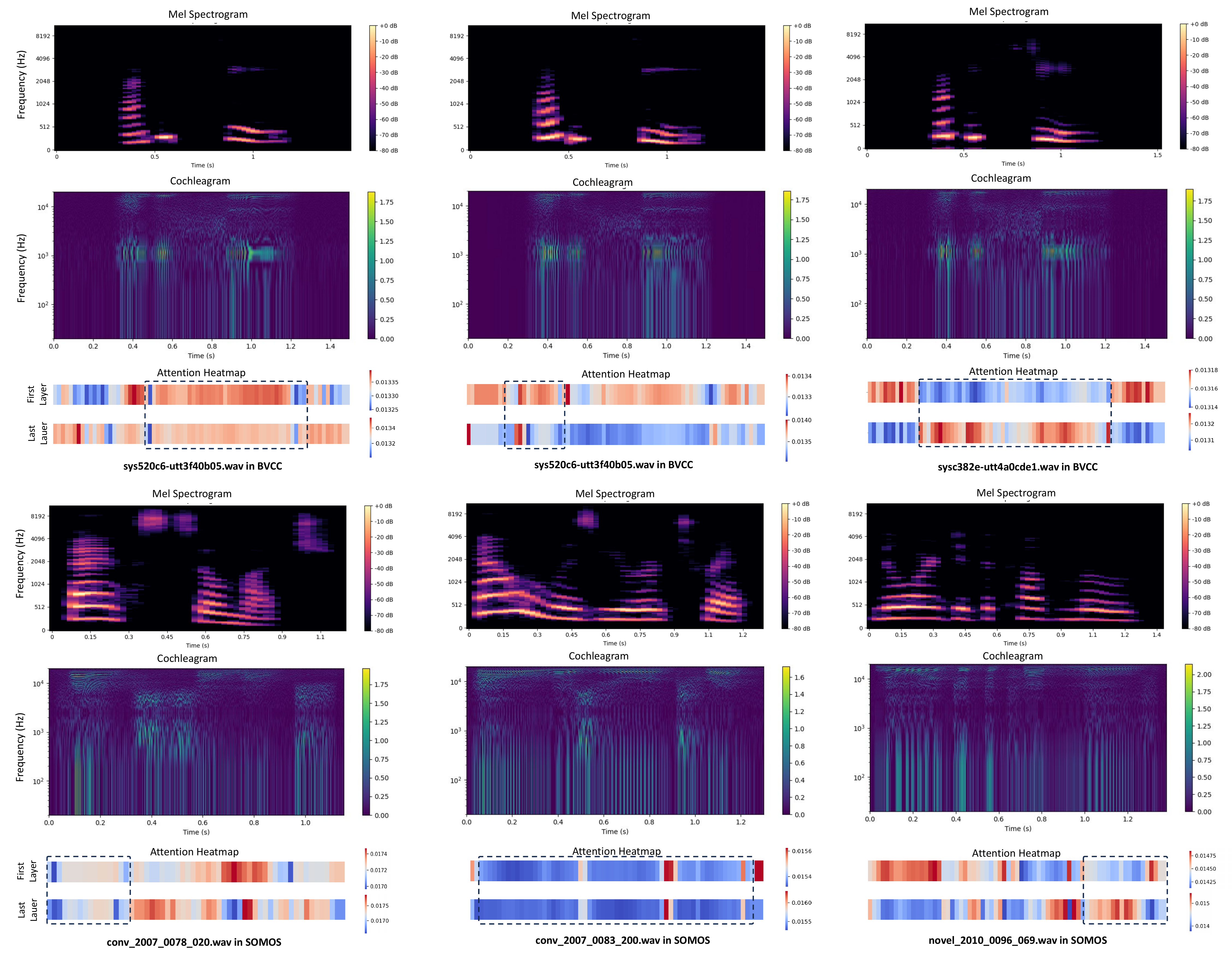} 
    \caption{Visualization of the Mel spectrograms, cochleagrams, and attention weight heatmaps for speech samples.}
    \label{fig:vis}
\end{figure*}

\section{Evaluation Metrics}
\label{eval}

Mean Squared Error (MSE) measures the error for MOS prediction:
\begin{equation}
    \text{MSE} = \frac{1}{n} \sum (y_i - \hat{y}_i)^2
\end{equation}
where \(n\) denotes the number of samples, with \(y\) representing the actual values and \(\hat{y}\) corresponding to the predicted values.

Linear correlation coefficient (LCC) compares bivariate data:
\begin{equation}
    \text{LCC} = \frac{n(\sum y\hat{y}) - (\sum y)(\sum \hat{y})}{\sqrt{\left[n\sum y^2 - (\sum y)^2\right]\left[n\sum \hat{y}^2 - (\sum \hat{y})^2\right]}}
\end{equation}
where \(\hat{y}\) denotes the predicted values and \(y\) represents the actual values.

Spearman rank correlation coefficient (SRCC) measures monotonic relationships:
\begin{equation}
    \text{SRCC} = 1 - \frac{6 \sum (r_i - \hat{r}_i)^2}{n(n^2 - 1)}
\end{equation}
where \(\hat{r}_i\) represents the ranks of predicted values and \(r_i\) signifies the ranks of true values, with \(n\) being the total count of samples.

Kendall tau (KTAU) measures ranked data consistency:
\begin{equation}
    \text{KTAU} = \frac{C - D}{C + D}
\end{equation}
where \(C\) is the number of concordant pairs and \(D\) is the number of discordant pairs.

\section{Visualization for Speech Samples}
\label{vis}

Here, we show more attention heatmap visualizations in Figure \ref{fig:vis}, which reveal how the cross-attention interactions of the first layer guide and influence deeper processing stages. The attention patterns of the first and last layers sometimes exhibit consistent similarities. For example, in the four left samples, parts of the attention weights in the final layer retain a distribution similar to that of the first layer. Furthermore, there are cases where the first and last layers do not maintain complete consistency, but the layers preserve similar regional segmentation, as shown in the two right samples. This shows the significant impact of auditory perception on semantic representation and reveals latent influence patterns. Such findings facilitate better modeling of the human process from auditory perception to semantic understanding and, ultimately, the judgment of speech quality.

\section{Introduction of Compared Models}
\label{compare}

Here, we provide a more detailed introduction to the compared models to clarify the techniques they employ.

\textbf{MOSNet\cite{Lo2019mosnet}} MOSNet develops a deep learning-based assessment model to predict human ratings of converted speech in voice conversion. The model architectures adopted include CNN, BLSTM, and a combined CNN-BLSTM, which use raw magnitude spectrograms as input and predict the mean opinion score (MOS) through fully connected layers and pooling operations.

\textbf{MBNet\cite{Leng2021MBNet}} MBNet uses individual judge scores in MOS datasets and handles score biases caused by personal preference. MBNet consists of a MeanNet convolution and BLSTM layers and a smaller BiasNet with convolution blocks to predict bias scores. The listener ID is needed as a prior dependency for the training, which is embedded and concatenated with the input spectrum in BiasNet.

\textbf{LDNet\cite{Huang2022LDnet}} LDNet addresses data scarcity and individual listener preference issues by directly predicting listener-wise perceived quality and also presents new inference methods for more stable and efficient prediction. LDNet is composed of an encoder that learns listener-independent features and a decoder that fuses listener information to generate listener-dependent features. 

\textbf{MOSA-Net\cite{Zezario2023MOSA-Net}} MOSA-Net can estimate speech quality, intelligibility, and distortion scores simultaneously. MOSA-Net adopts a structure that combines a CNN and a BLSTM for feature extraction. Latent representations of wav2vec2 and HuBERT are used as inputs to MOSA-Net to capture context information.

\textbf{SSL-MOS\cite{Cooper2022Generalization}} SSL-MOS fine-tunes wav2vec2 models by mean-pooling the output embeddings, adding a linear output layer, and training with L1 loss.

\textbf{RAMP\cite{wang2023RAMP}} RAMP is a retrieval-augmented MOS prediction method, which uses a confidence-based dynamic weighting scheme to enhance the decoder in SSL-based frameworks. RAMP(P) is the parametric path in RAMP, which is based on a neural network multitask architecture and predicts MOS scores and confidence levels through regression and classification heads. RAMP(NP) is the non-parametric path of RAMP, which uses the kNN model to predict MOS scores based on feature retrieval. RAMP combines the results of RAMP(P) and RAMP(NP) and improves prediction performance by dynamically adjusting the retrieval scope and fusion weights.

\textbf{Vioni et al.\cite{Voini2023Investigating}} The model incorporates prosodic and linguistic features as additional input for MOS prediction. The SSL model Wav2vec2 is used, which provides pre-trained speech representations to enhance generalization in MOS prediction, and a BERT model is used for linguistic encoding. The model architectures, including MOSNet, LDNet, and SSL-MOS, are augmented with feature encoders (e.g., BiGRU, BLSTM, feed-forward networks) to integrate additional prosodic and linguistic inputs alongside spectral or speech signal features.

\textbf{Wang et al.\cite{Wang2025Enabling}} The model uses auditory large language models (LLM) for automatic speech quality assessment, which are fine-tuned via task-specific prompts to predict MOS and generate natural-language-based evaluations. Auditory LLMs typically consist of an audio encoder, a connection module, and an LLM. The audio encoder extracts features, the connection module aligns the feature space, and the LLM generates responses based on the input audio and text prompt.

\textbf{DDOS\cite{tseng2022DDOS}} DDOS uses domain-adaptive pre-training and models opinion score distribution to enhance generalization. DDOS consists of Wav2vec2 for encoding, a base MOS predictor with regression and distribution heads, and a refinement layer, and the model is trained in three stages. DDOS also adopts listener modeling to capture listener bias.

\textbf{UTMOS\cite{saeki2022utmos}} UTMOS uses ensemble learning of strong and weak learners. The strong learner in UTMOS is an SSL-based neural network model that directly takes a speech waveform as input, with an architecture incorporating a fine-tuned SSL model, contrastive loss, listener and data-domain dependent learning, phoneme encoding, and data augmentation. The weak learner in UTMOS is a combination of basic machine-learning regression models, which use utterance-level features extracted from pretrained SSL models as input to predict MOS scores.

\bibliographystyle{ACM-Reference-Format}
\bibliography{reference}